\documentclass[%
 reprint,
 amsmath,amssymb,
 aps,
]{revtex4-2}

\usepackage{graphicx}
\usepackage{color}%
\usepackage{dcolumn}
\usepackage{bm}
\usepackage{braket}

\begin{document}

\preprint{APS/123-QED}

\title{
Catastrophic failure of quantum annealing owing
to
non-stoquastic Hamiltonian and its avoidance by decoherence
}

\author{Takashi Imoto}

\affiliation{%
 Research Center for Emerging Computing Technologies (RCECT), National Institute of Advanced Industrial Science and Technology (AIST),
1-1-1 Umezono, Tsukuba, Ibaraki 305-8568, Japan.
}
\author{Yuichiro Matsuzaki}
\email{matsuzaki.yuichiro@aist.go.jp}
\affiliation{Research Center for Emerging Computing Technologies, National Institute of Advanced Industrial Science and Technology (AIST), Umezono1-1-1, Tsukuba, Ibaraki 305-8568, Japan.}
\affiliation{NEC-AIST Quantum Technology Cooperative Research Laboratory, National Institute of Advanced Industrial Science and Technology (AIST), Tsukuba, Ibaraki 305-8568, Japan}

\date{\today}

\begin{abstract}
Quantum annealing (QA) is a promising method for solving combinatorial optimization problems whose solutions are embedded into a ground state of the Ising Hamiltonian.
This method employs two types of Hamiltonians:
a driver Hamiltonian and a problem Hamiltonian. After a sufficiently slow change from the driver Hamiltonian to the problem Hamiltonian, we can obtain the target ground state that corresponds to the solution.
The inclusion of non-stoquastic terms in the driver Hamiltonian is believed to enhance the efficiency of the QA.
Meanwhile, decoherence is regarded as of the main obstacles for QA.
Here, we present examples showing that non-stoaquastic Hamiltonians can lead to catastrophic failure of QA, whereas a certain decoherence process can be used to avoid such failure.
More specifically, when we include anti-ferromagnetic interactions (i.e., typical non-stoquastic terms) in the Hamiltonian, we are unable to prepare the target ground state even with an infinitely long annealing time for some specific cases.
In our example, owing to a symmetry, the Hamiltonian 
is block-diagonalized, and a crossing occurs during the QA, which leads to a complete failure of the ground-state search. 
Moreover, we show that, when we add a certain type of decoherence, we can obtain the ground state after QA for these cases.
This is because, even when symmetry exists in isolated quantum systems, the environment breaks the symmetry.
Our counter intuitive results provide a deep insight into the fundamental mechanism of QA.

\begin{description}
\item[Usage]
Secondary publications and information retrieval purposes.
\item[Structure]
You may use the \texttt{description} environment to structure your abstract;
use the optional argument of the \verb+\item+ command to give the category of each item. 
\end{description}
\end{abstract}

\maketitle


\section{Introduction}

Recently, quantum annealing (QA) has attracted considerable attention owing to its potential applications for solving practical problems.
QA is expected to not only solve combinatorial optimization problems but also simulate quantum many-body systems\cite{kadowaki1998quantum,farhi2000quantum,farhi2001quantum}.
In QA, after we prepare a ground state of a driver Hamiltonian, we gradually change the Hamiltonian from the driver Hamiltonian to the problem Hamiltonian. Furthermore, an adiabatic theorem guarantees that we can obtain a ground state of the problem Hamiltonian with QA as long as the dynamics is adiabatic.

There are two types of applications of QA.
One application is to find a ground state of the Ising Hamiltonian\cite{schrijver2005history}.
A combinational optimization problem can be mapped into a ground state search of the Ising Hamiltonian.
Moreover, some types of clustering and machine learning can be conducted using this type of QA.
Efficient clustering using QA has been reported
\cite{kumar2018quantum, kurihara2014quantum}.
In addition, machine learning using QA has been proposed
\cite{kumar2018quantum, kurihara2014quantum,adachi2015application, wilson2021quantum, li2020limitations, sasdelli2021quantum, neven2008training, neven2012qboost, willsch2020support, winci2020path}.
Furthermore, QA has been applied to topological data analysis(TDA)\cite{berwald2018computing}.
The other application is to find a ground state of the problem Hamiltonian including non-zero off-diagonal terms.
To understand the condensed matter physics, it is crucial to investigate a correlation function of the ground state, and QA is useful for such a study.
In addition, applications of QA to quantum chemistry have been reported \cite{bravyi2002fermionic, seeley2012bravyi, tranter2015b, babbush2014adiabatic, xia2017electronic, seki2021excited}.

Several devices have been developed for performing QA.
D-Wave Systems Inc. developed a device for performing QA using thousands of superconducting flux qubits~\cite{johnson2011quantum, barends2016digitized, harris2018phase}.
Many demonstrations using this device for QA have been reported~\cite{adachi2015application, hu2019quantum, joseph2021two, king2018observation}.
A Kerr-nonlinear parametric oscillator(KPO) is another device for performing QA~\cite{goto2016bifurcation,puri2017quantum,wang2019quantum,grimm2020stabilization,yamaji2022spectroscopic}.
Some methods use capacitively shunted flux qubits with a long coherence time
for QA~\cite{matsuzaki2020quantum, imoto2022obtaining}.

Decoherence is one of the main obstacles for QA \cite{albash2015decoherence}.
Owing to unwanted coupling with the environment, decoherence should be considered during QA. In particular, thermal excitation can lead to a decrease in the success probability of QA.

Non-adiabatic transitions also pose problems for QA.
For example, when the first-order quantum phase transition occurs, the energy gap becomes exponentially smaller as the system size increases, and it becomes difficult to satisfy the adiabacity condition in QA.
A promising approach for avoiding the first-order quantum phase transition is to add a non-stoquastic Hamiltonian to the driver Hamiltonian.
By adding anti-ferromagnetic interactions (which are non-stoquastic) to the Hamiltonian, we can avoid the first-order phase transition for some specific problem Hamiltonians~\cite{seki2012quantum, seki2015quantum,susa2022nonsto}. 

\begin{figure}[ht]
    \includegraphics[width=90mm]{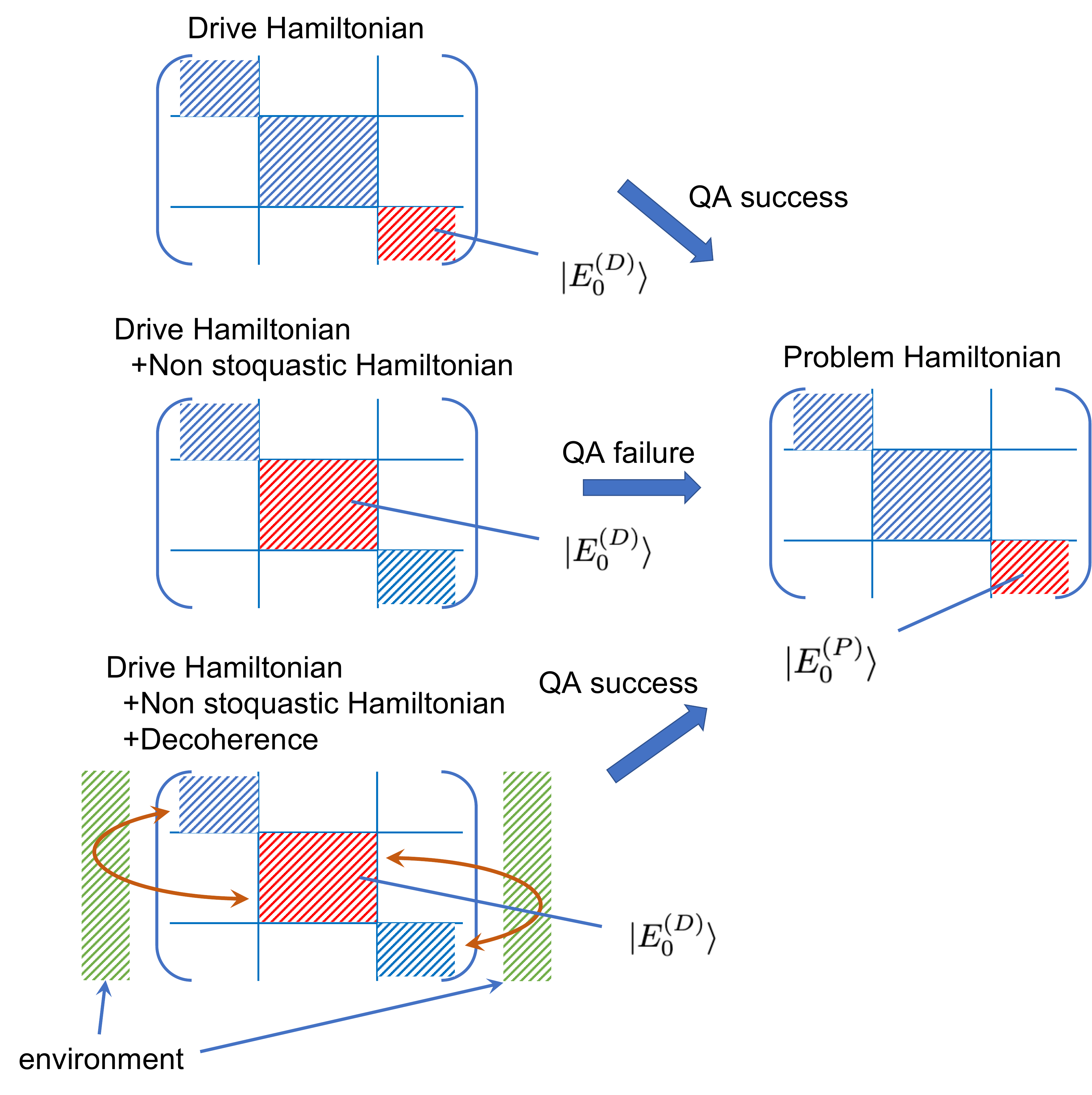}
    \caption{
    Structure of the annealing Hamiltonian investigated in the present paper. The annealing Hamiltonian can be bock-diagonalized when there is symmetry.
    For our choice of the problem Hamiltonian, when the driver Hamiltonian is the transverse magnetic field, the ground state of the driver Hamiltonian belongs to the same sector as that of the problem Hamiltonian.
    Meanwhile, if we add a non-stoquastic Hamiltonian to the transverse-field driver Hamiltonian, the sector of the ground state of the driver Hamiltonian becomes different from that of the problem Hamiltonian.
    Moreover, when a certain type of decoherence breaks the symmetry, we can induce transitions from the ground state of the driver Hamiltonian to that of the problem Hamiltonian.
    }
    \label{fig:concept_this_paper}
\end{figure}

In this paper, we challenge the common wisdom that non-stoquastic Hamiltonians improve the performance of QA whereas decoherence degrades the performance of QA. More specifically, we present counter-intuitive examples showing that non-stoquastic Hamiltonians can cause catastrophic failure of QA in the sense that we cannot obtain a ground state even for an infinitely long annealing time, and a certain type of decoherence can be used to recover the performance of QA in such cases.
In our examples, we consider either an Ising Hamiltonian or an XXZ model for the problem Hamiltonian and transverse fields with anti-ferromagnetic interactions for the driver Hamiltonian.
The driver Hamiltonian and problem Hamiltonian have a common symmetry in that both of these Hamiltonians commute with a certain observable.
In this case, the Hamiltonian can be block diagonalized (Fig \ref{fig:concept_this_paper}).
If the coupling strength of the anti-ferromagnetic interactions is smaller than a certain threshold, the ground state of the driver Hamiltonian and that of the problem Hamiltonian belong to the same sector, and successful QA can be achieved if the annealing time is sufficiently long. However,
if the coupling strength of the anti-ferromagnetic interactions is larger than the threshold, the ground state of the driver Hamiltonian and that of the problem Hamiltonian belong to different sectors. In this case, the ground-state search with QA completely fails in the sense that the fidelity with the ground state becomes 0, because the transitions between different sectors are prohibited owing to the block-diagonal structure of the Hamiltonian.
Even for such cases, we show that, if we add a certain type of decoherence to break the symmetry, we can obtain the ground state with QA. In addition, we confirm that, when the environmental temperature is sufficiently low, the success probability becomes nearly unity with decoherence.
Thus, our results challenge the common wisdom about non-stoquastic Hamiltonians and decoherence, thereby providing a deeper understanding of QA.

The remainder of this paper is organized as follows.
Section II reviews QA as well as the relation between symmetry and QA.
Section III discusses the main results.
Finally, Section IV summarizes our findings.

\section{Overview}
In this section, we review the QA as well as the relation between symmetry and QA..

\subsection{Review of QA}
We review QA for the ground-state search~\cite{kadowaki1998quantum, farhi2000quantum, farhi2001quantum}.
In QA, there are two types of Hamiltonians: a driver Hamiltonian and a problem Hamiltonian. We use the driver Hamiltonian to induce quantum fluctuations during QA, and we aim to find a ground state of the problem Hamiltonian.
We define the annealing Hamiltonian using the driver Hamiltonian and the problem Hamiltonian as follows:

\begin{align}
    H(t)\equiv\biggl(1-\frac{t}{T}\biggr)H_{D}+\frac{t}{T}H_{P}
\end{align}
where $H_{D}$ denotes the drive Hamiltonian, $H_{P}$ denotes the problem Hamiltonian, $t$ denotes the time, and $T$ denotes the annealing time.
In QA, we prepare a ground state of the drive Hamiltonian at $t=0$,
and we gradually change the Hamiltonian from the driver Hamiltonian to the problem Hamiltonian. As long as the dynamics is adiabatic, the ground state of the problem Hamiltonian is prepared at $t=T$.

There is a known condition for satisfying the adiabacity condition during QA.
The adiabatic condition is given by
\begin{align}
    \frac{\bra{j(t)}\partial_{t}H(t)\ket{0(t)}}{\Delta_{j}(t)^{2}} \ll 1
\end{align}
where $\Delta_{j}(t)$ denote the energy gap between the ground state and the $j$-th excited state at the time $t$, $\ket{j(t)}$ is the $j$-th excited state, 
and
$\ket{0(t)}$ is the ground state~\cite{kato1950adiabatic, messiah2014quantum, morita2008mathematical}. 
It is possible to evaluate this adiabatic condition in an experiment
\cite{matsuzaki2021direct,russo2021evaluating,schiffer2022adiabatic,mori2022evaluate}.

Thus, when the minimum energy gap becomes exponentially smaller as the number of qubits increases, which corresponds to the first-order quantum phase transition, the annealing time should be exponentially large in the number of qubits.
If the dynamics is not adiabatic during QA, some population of the ground state is transferred to that of the excited states.
This is referred to as a non-adiabatic transition.

Several attempts have been made to avoid the first-order phase transition during QA.
A promising approach is to include non-stoquastic terms in the driver Hamiltonian.
A Hamiltonian is said to be stoquastic if all the off-diagonal matrix elements are real and non-positive in a given basis.
It is known that, when one uses a transverse-field driver Hamiltonian (that is stoquastic) to obtain a ground state of the p-spin model,
the first-order phase transition occurs during QA\cite{jorg2010energy}.
In this case, it has been shown that, for the driver Hamiltonian, if we add anti-ferromagnetic interactions that are non-stoquastic, then we can avoid the first-order quantum phase transition \cite{seki2012quantum}.
A similar conclusion has been drawn for the case of the Hopfield model as the problem Hamiltonian \cite{seki2015quantum}.
Moreover, an inhomogeneous driver Hamiltonian helps avoid the first-order phase transition for QA to obtain a ground state of the p-spin model~\cite{susa2018exponential, susa2018quantum}.
Even when we perform bifurcation-based quantum annealing with spin-1 particles \cite{takahashi2022bifurcation},
the non-stoquastic Hamiltonian is useful to obtain a ground state of the p-spin model by avoiding the first-order phase transition~\cite{susa2022nonsto}.

For actual devices, we must consider the effect of decoherence from the environment.
Studies have investigated the existence of competition between non-adiabatic transitions and decoherence\cite{keck2017dissipation, novotny2016quantum}.
In addition, some cases of improved QA accuracy have been reported under strong decoherence\cite{passarelli2018dissipative}.
There are other methods to suppress decoherence
using variational techniques \cite{susa2021variational, matsuura2021variationally,imoto2021quantum}.

\subsection{Review of the relation between a symmetry and QA}\label{sec:review_symm}
We review how the symmetry of a system can impair the performance of QA.
Suppose that there exists an observable 
$K$ that commutes with the annealing Hamiltonian at any time ,i.e.,
\begin{align}
[K, H(t)] = 0\ \ (0\leq\forall t\leq T).    
\end{align}
In this case, the annealing Hamiltonian can be block diagonalized into sectors by applying a suitable unitary operator. 
The operator $K$ is referred to as a conservation quantity.
Thus, transitions between different sectors are prohibited during the QA\cite{imoto2022quantum}. 
When the sector of the ground state of the drive Hamiltonian is different from that of the ground state of the problem Hamiltonian, i.e.
\begin{align}
    \bra{\mbox{gs}(t=0)}K\ket{\mbox{gs}(t=0)}\neq\bra{\mbox{gs}(t=T)}K\ket{\mbox{gs}(t=T)}
\end{align}
where $\ket{\mbox{gs}(t)}$ denotes the ground state of the annealing Hamiltonian at $t$, QA fails
\cite{imoto2022obtaining, francis2022determining}.
Concretely, the fidelity between the exact ground state of the problem Hamiltonian and the state obtained by QA become zero even if an infinitely long annealing time is considered.

\section{Main result}
In this section, we present two main results. First, we analyze QA with a non-stoquastic Hamiltonian, and we present an example showing that, owing to the inclusion of non-stoquastic terms, QA can fail to find a ground state even after an infinitely long annealing time.
Second, we investigate the effect of decoherence during QA for such a case, and we show that a certain decoherence process recovers the ability to find the ground state.

\subsection{Failure of QA owing to a non-stoquastic Hamiltonian}

In general, a stoquastic Hamiltonian is defined as an operator having only non-positive (or only non-negative) off-diagonal elements in the computational basis~\cite{albash2018adiabatic}.
By contrast, a non-stoquastic Hamiltonian has both positive and negative off-diagonal elements in the computational basis.
A typical example of a non-stoquastic Hamiltonian is the XX anti-ferromagnetic interaction, which can improve the accuracy of QA for specific cases
\cite{seki2012quantum, seki2015quantum}.
This anti-ferromagnetic interaction Hamiltonian is defined as
\begin{align}
    H_{XX}=N\biggl(\frac{1}{N}\sum_{i}^{N}\hat{\sigma}_{i}^{x}\biggr)\label{eq:xx_non-stoquastic}
\end{align}
where $N$ denote the number of qubits, and $\hat{\sigma}_{i}^{a}, (a=x,y,z)$ denote the the Pauli matrices defined on the $i$-th site.
In this study, we focus on the XX anti-ferromagnetic interaction as the non-stoquastic Hamiltonian.

Here, we define the annealing Hamiltonian with the non-stoquastic Hamiltonian as
\begin{align}
    H^{(NS)}(t, \alpha)\equiv\biggl(1-\frac{t}{T}\biggr)(H_{D}+\alpha H_{XX})+\frac{t}{T}H_{P}.\label{hamiltonianwithns}
\end{align}

In this paper, we consider the case where $H^{(NS)}(t, \alpha)$ has a conserved observable $K$ for any time $t$ and any parameters $\alpha$.
In other words,
\begin{align}
\exists\  K\  s.t\  [K, H^{(NS)}(t, \alpha)] = 0\ \ (0\leq\forall t\leq T, \forall \alpha).    
\end{align}

Therefore, the ground state of the Hamiltonian is also an eigenstate of $K$.
We present an example showing that, owing to the inclusion of non-stoquastic terms, QA can fail to find a ground state.
Thus, there exists a case such that the sector of the ground states satisfies

\begin{align}
    \bra{gs(t=0, \alpha=0)}&K\ket{gs(t=0, \alpha=0)}\notag\\
    &=\bra{gs(t=T)}K\ket{gs(t=T)}
\end{align}
and 
\begin{align}
    \exists\  \alpha_{0}\ s.t. \bra{gs(t=0, \alpha=\alpha_{0})}&K\ket{gs(t=0, \alpha=\alpha_{0})}\notag\\
    &\neq\bra{gs(t=T)}K\ket{gs(t=T)}
\end{align}
where $\ket{gs(t, \alpha)}$ denotes the ground state of the annealing Hamiltonian with the non-stoquastic term $H^{(NS)}(t, \alpha)$.
In these examples, the problem Hamiltonian is either a fully connected Ising model or an XXZ spin chain.
Further details are discussed in Sections \ref{sec:example_ising} and \ref{sec:example_xxz}.

\subsection{Avoidance of failure caused by symmetry using decoherence}

In this subsection, we show that, by adding a certain decoherence process,
we can avoid the problem of QA failure due to symmetry.
In previous calculations, we assumed that quantum systems are closed to any environment.
Even if a conserved quantity exists in an isolated quantum system, a perturbation from the environment can break such symmetry.
Indeed, decoherence can not only break the symmetry but also induce unwanted transitions to excited states.
To avoid the thermal excitation, the environmental temperature should be sufficiently low.
In previous studies, the Gorini--Kossakowski--Sudarshan--Lindblad (GKSL) master equations~\cite{gorini1976completely, lindblad1976generators} were occasionally employed to consider decoherence, where the Lindblad operators were selected as $\hat{\sigma}_{\pm}$
phenomenologically. However, in this case, there is no information about the system Hamiltonian in the Lindblad operator. Therefore, the Lindblad operator induces a transition to the excited states of the system, even at a low temperature in our case (see Appendix \ref{sec:numerical_gksl}).
The problem arises from the fact that the Lindblad operator $\hat{\sigma}_{\pm}$ is derived when there is no interaction between qubits. This means that, if there are non-negligible interactions between qubits, we cannot employ $\hat{\sigma}_{\pm}$ as the Lindblad operator for the GKSL master equation to describe the energy relaxation.
Therefore, to consider more realistic situations, we employ the quantum adiabatic Markovian master equations 
~\cite{redfield1957theory, redfield1965theory, albash2012quantum}, where the noise operator is derived from the first-principles calculation by using a microscopic model.
We explain the master equation in detail in the next subsection.

\subsection{Quantum adiabatic Markovian master equations without the rotating wave approximation}

Here, we introduce the quantum adiabatic Markovian master equations\cite{redfield1957theory, redfield1965theory, rivas2012open, breuer2002theory,albash2012quantum} that we employ in this paper.
Suppose that the total Hamiltonian for the system and the environment is given by
\begin{align}
    H=H_{sys}+H_{bath}+H_{Int}
\end{align}
where $H_{sys}$, $H_{bath}$, and, $H_{I}$ are the system, bath, and interaction Hamiltonian, respectively.
Here, the interaction Hamiltonian can be expressed as 
$H_{I}=\sum_{k=1}^{M}A_{k}\otimes B_{k}$ where $A_{\alpha}$ denotes the noise operator acting on the system and $B_{\alpha}$ denotes the operator acting on the environment.
The dynamics of the total Hamiltonian is described by the von Neumann equation:
\begin{align}
    \frac{d}{dt}\rho(t)=-i[H, \rho].
\end{align}

By assuming the Born-Markov approximation and small non-adiabatic transitions, the quantum adiabatic Markovian master equations are given by \cite{albash2012quantum}
\begin{align}
    \frac{d}{dt}&\rho(t)=\sum_{k,l}\sum_{\omega, \omega'}e^{i(\omega-\omega')}\Gamma_{kl}(\omega')\notag\\
    &\times\biggl\{A_{l}(\omega')\rho(t)A_{k}^{\dag}(\omega)-A_{k}^{\dag}(\omega)A_{l}(\omega')\rho(t)\biggr\}+h.c
    \label{eq:redfield_eq}
\end{align}
where $\rho(t)$ denotes the density matrix, $A_{k}(\omega)=\sum_{\epsilon'-\epsilon=\omega}\ket{\psi_{\epsilon}}\bra{\psi_{\epsilon}}A_{k}\ket{\psi_{\epsilon'}}\bra{\psi_{\epsilon'}}$, $\epsilon$ denotes the eigenvalue of the Hamiltonian $H(t)$, $\ket{\psi_{\epsilon}}$ denotes the eigenvector of the Hamiltonian $H(t)$, $\omega$ denotes the energy difference, and $\Gamma_{kl}$ denotes the power spectrum density.
Throughout this paper, we set $M=1$; hence,
we denote $A_{k}$ simply as $A$.
We select the noise operator as $A=\sum_{j=1}^{N}\hat{\sigma}^{(y)}_{j}$, and select an Ohmic spectral density given by
\begin{align}
    \Gamma(\omega)= 
    \begin{cases}
        {\eta\omega\biggl(\frac{1}{e^{\omega/T_{env}}-1}+1\biggr) \ (\omega > 0)}\\
        {\eta T \ (\omega = 0)}\\
        {\eta(-\omega)\biggl(\frac{1}{e^{-\omega/T_{env}}-1}+1\biggr) \ (\omega < 0)}
    \end{cases}
\end{align}
where $\eta$ denote the strength of decoherence and $T_{env}$ denote the temperature of the environment.
When we conduct numerical simulations,
we introduce the cut-off parameters $\omega_{c}$ and $\epsilon$.
The resulting power spectrum density is given by
\begin{align}
    \Gamma(\omega)^{(co)}= 
    \begin{cases}
        {\eta\omega e^{-\frac{w}{w_{c}}}\biggl(\frac{1}{e^{\omega/T_{env}}-1+\epsilon}+1\biggr) \ (\omega > 0)}\\
        {\eta T_{env} \ (\omega = 0)}\\
        {\eta(-\omega)e^{\frac{w}{w_{c}}}\biggl(\frac{1}{e^{-\omega/T_{env}}-1+\epsilon}+1\biggr) \ (\omega < 0)}
    \end{cases}
\end{align}

In this paper, we use this power spectrum density and we set $\omega_{c}=20$, $\epsilon=10^{-7}$, and $\eta=0.1$.

\subsection{Example 1: Ising problem Hamiltonian}\label{sec:example_ising}

\begin{figure}[h!]
    \includegraphics[width=80mm]{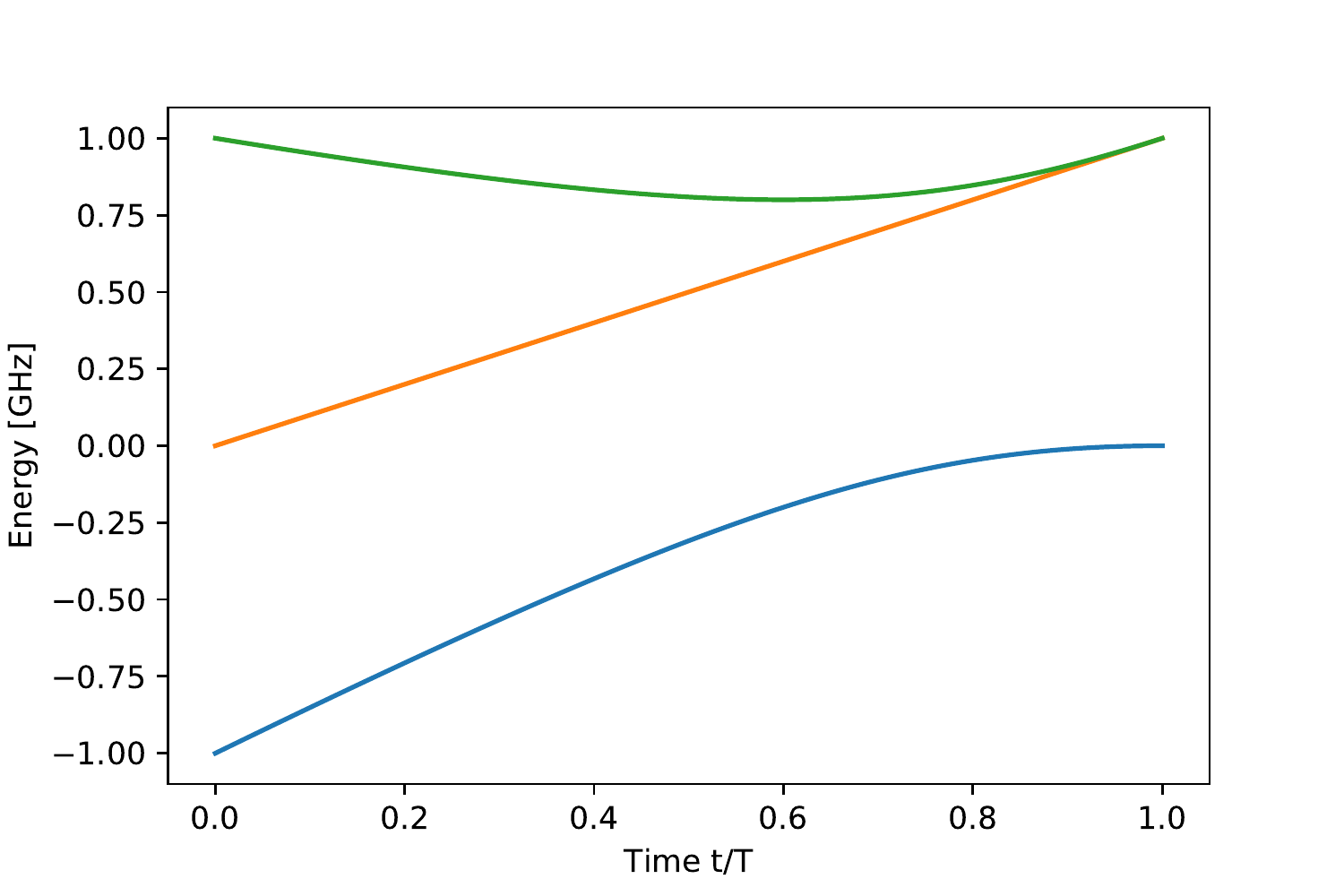}
    \caption{For the Ising model as the problem Hamiltonian,
    we plot the energy spectrum of the annealing Hamiltonian against time $t$. Here, the driver Hamiltonian is the transverse field and the problem Hamiltonian is the fully connected Ising model. There is no level crossing in this energy diagram.
    }
    \label{fig:ising_no_non-stoquastic_energy}
\end{figure}

\begin{figure}[h!]
    \includegraphics[width=80mm]{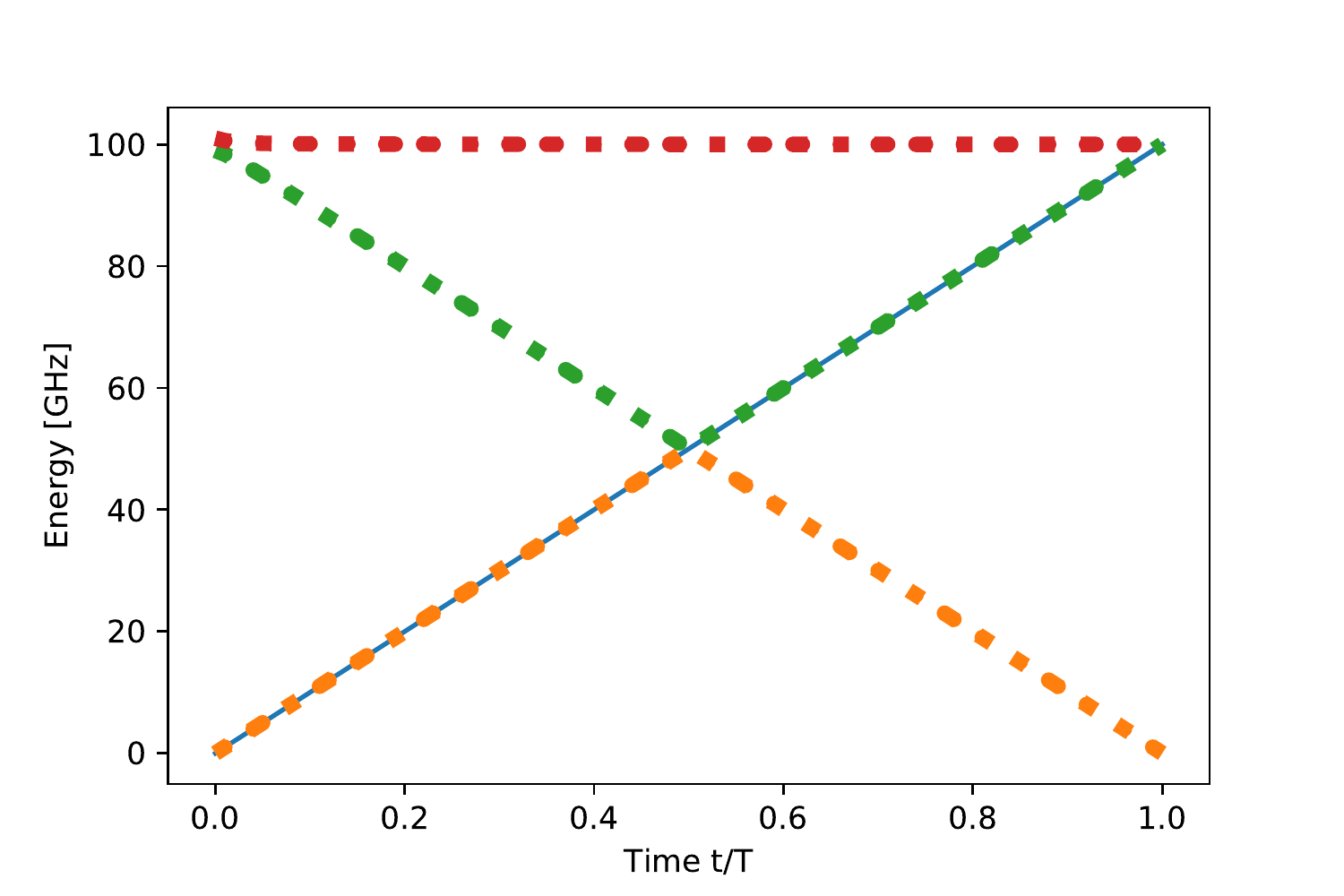}
    \caption{
    With the Ising model as the problem Hamiltonian, where we employ the driver Hamiltonian with the non-stoquastic term, we solve the time-dependent Schrodinger equation and plot the expectation values of the Hamiltonian during QA against $t/T$ by using a continuous line.
    The dotted line represents the energy spectrum during QA.
    This plot shows the existence of the level crossing. Here, we set $\alpha =100$ and $T=1000$.
    }
    \label{fig:ising_with_non-stoquastic_energy}
\end{figure}

\begin{figure}[h!]
    \includegraphics[width=80mm]{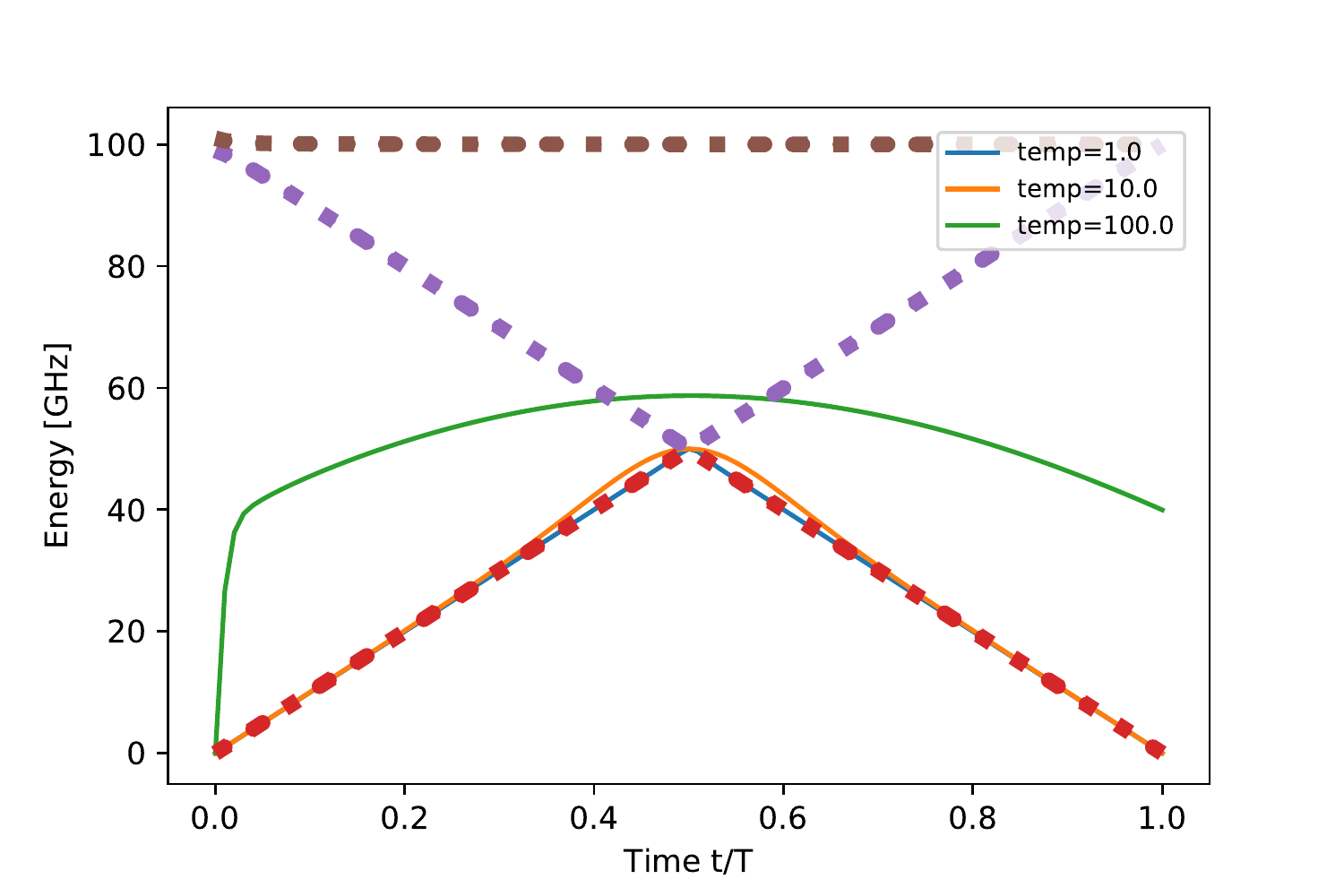}
    \caption{With the Ising model as the problem Hamiltonian,
    we plot the energy against $t/T$ via QA for the temperatures of $1.0, 10.0, 100.0$ by using continuous lines.
    The dotted line represents the energy spectrum against each time $t/T$.
    QA succeeds in preparing the ground state with high accuracy if the temperature of the environment is sufficiently low.
    We set $\alpha=100$ and $T=1000$.
    }
    \label{fig:ising_with_non-stoquastic_energy_with_decoherence}
\end{figure}
As the first example, we consider the case in which the problem Hamiltonian is the fully connected Ising model.
When the driver Hamiltonian is the transverse field without the non-stoquatic term, the Hamiltonians are given by 
\begin{align}
    H_{D}&=\sum_{i=1}^{N}\hat{\sigma}_{i}^{x}\label{nononsd}\\
    H_{P}&=N\biggl(\frac{1}{N}\sum_{i=1}^{N}\hat{\sigma}_{i}^{z}\biggr)^{2} \label{nononsp},
\end{align}
where $N$ is the number of qubits.
In this case, the annealing Hamiltonian $H(t)\equiv (1-t/T)H_{D}+(t/T)H_{P}$ has a conserved quantity $K=e^{i\frac{\pi}{2}\sum_{i=1}^{N}\hat{\sigma}_{i}^{(x)}}$.
In other words, the operator $K$ satisfies the following condition:
\begin{align}
[H(t), K]=0.
\end{align}
for all $t$.
Here, the sector of the ground state of $H_{D}$ is the same as that of $H_{P}$. This is analytically demonstrated in Appendix \ref{sec:appendix_symm}.

For $N=2$, we plot an energy diagram of the annealing Hamiltonian with Eqs (\ref{nononsd}) and  (\ref{nononsp})
as shown in Fig. \ref{fig:ising_no_non-stoquastic_energy}.
Throughout this paper, as the state is confined in the maximum angular momentum, we conduct numerical simulations using the Dicke basis \cite{dicke1954coherence}.
As expected, the crossing does not occur in the energy diagram; hence, we can find the ground state with QA as long as the dynamics is adiabatic.

Meanwhile, let us consider the annealing Hamiltonian with a non-stoquastic term
$H^{(NS)}(t, \alpha)= (1-t/T)(H_{D}+\alpha H_{XX})+(t/T)H_{P}$. The quantity $K$ is conserved for this Hamiltonian, and we have $[H^{(NS)}(t, \alpha), K]=0$.
Moreover, in this case, the sector of the ground state of the driver Hamiltonian is different from that of the problem Hamiltonian (see Appendix \ref{sec:appendix_symm}). This means that the ground-state search with QA fails even if we consider an infinitely long annealing time.

For $N=2$, we plot the energy diagram of the annealing Hamiltonian with the non-stoquastic term, as shown in Fig. \ref{fig:ising_with_non-stoquastic_energy}. This result shows that there is a level crossing during QA.
Furthermore, we solve the time-dependent Schrodinger equation and plot the expectation value of the Hamiltonian during QA, as shown in Fig. \ref{fig:ising_with_non-stoquastic_energy}. Again, we confirm that the crossing occurs during QA.

Next, we propose a method to avoid failure of QA using decoherence.
We consider a noise model to break the symmetry of the Hamiltonian.
More specifically, we select a noise operator to satisfy $[A, K]\neq 0$.
In this case, we expect that, owing to noise-induced symmetry breaking, there should be a transition from the ground state of the driver Hamiltonian to that of the problem Hamiltonian.

We conduct the numerical calculation using the quantum adiabatic Markovian master equations 
(\ref{eq:redfield_eq}).
We plot the energy spectrum and the energy expectation value $\langle H\rangle ={\rm{Tr}}[H(t)\rho(t)]$, as shown in Fig. \ref{fig:ising_with_non-stoquastic_energy_with_decoherence}.
This figure shows that the ground state is obtained by QA with high probability when the temperature is sufficiently low.
We can interpret this result as follows.
Owing to the symmetry of the Hamiltonian, the system will be prepared in the excited state
if the decoherence is negligible. However, owing to the existence of the decoherence, an energy relaxation from the excited state to the ground state occurs, and the system is prepared in a ground state as long as the temperature is sufficiently low.
Furthermore, we conduct numerical simulations in the case of many qubits (see Appendix \ref{manyqubits}), and we draw the same conclusion.

\subsection{Example 2: XXZ problem Hamiltonian}\label{sec:example_xxz}
As the second example, we consider the case in which the problem Hamiltonian is a fully connected XXZ spin chain.
The problem Hamiltonian is given by
\begin{align}
    H_{P}^{(XXZ)}=N\biggl(\frac{1}{N}\sum_{i=1}^{N}\hat{\sigma}_{i}^{x}\biggr)^{2}&+N\biggl(\frac{1}{N}\sum_{i=1}^{N}\hat{\sigma}_{i}^{y}\biggr)^{2}\notag\\
    &+\Delta N\biggl(\frac{1}{N}\sum_{i=1}^{N}\hat{\sigma}_{i}^{z}\biggr)^{2}\label{eq:xxz}
\end{align}
where $N$ is the number of qubits.
We select the transverse field as the driver Hamiltonian described in Eq. (\ref{nononsd}).
In this case, again, the operator $K=e^{i\frac{\pi}{2}\sum_{i=1}^{N}\hat{\sigma}_{i}^{(x)}}$ is the conserved quantity of the annealing Hamiltonian.
Throughout this paper, we set $\Delta=1.5$ and $N=2$.

\begin{figure}[ht]
    \includegraphics[width=80mm]{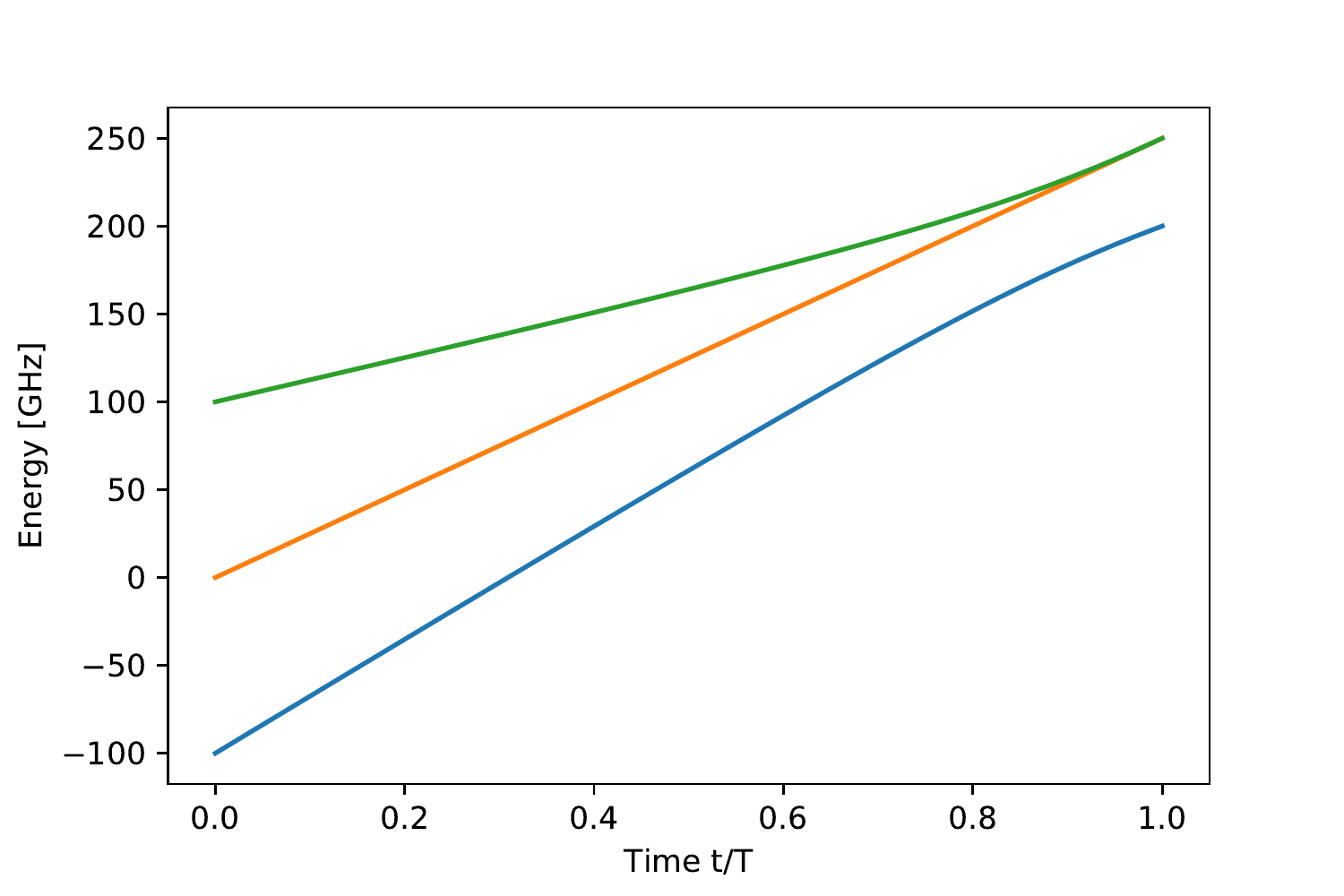}
    \caption{With the XXZ model as the problem Hamiltonian,
    we plot the energy spectrum of the annealing Hamiltonian without the non-stoquastic Hamiltonian against each time $t/T$.
    We note that the anisotropic parameter $\Delta=1.5$.
    We can see that the crossing does not occur.
    Here, we set $T =1000$.
    }
    \label{fig:xxz_no_non-stoquastic_energy}
\end{figure}

We plot an energy diagram of the annealing Hamiltonian with Eqs. (\ref{eq:xxz}) and (\ref{nononsd}), as shown in Fig. \ref{fig:xxz_no_non-stoquastic_energy}. There is no level crossing in the energy diagram.

\begin{figure}[ht]
    \includegraphics[width=80mm]{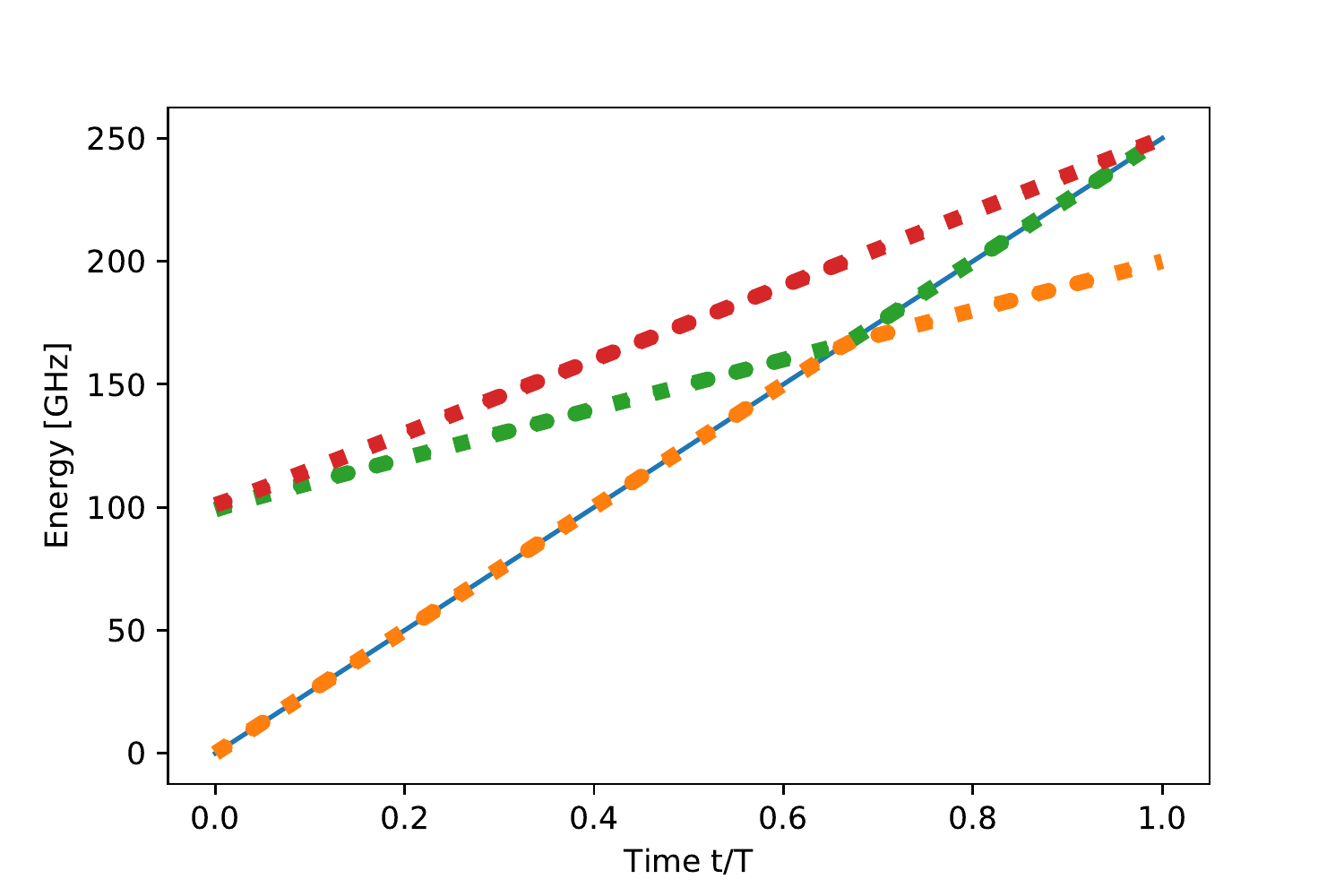}
    \caption{
    With the XXZ model as the problem Hamiltonian, we plot the energy during QA against $t/T$ by using a continuous line,
    where we employ the driver Hamiltonian with the non-stoquastic term.  The dotted line represents the energy spectrum during QA.
    We note that the anisotropic parameter is set as $\Delta=1.5$.
    In addition, we set $\alpha=100$ and $T=1000$.
    We can see that the crossing occurs.
    }
    \label{fig:xxz_with_non-stoquastic_energy}
\end{figure}

Let us consider the annealing Hamiltonian with a non-stoquastic term
$H^{(NS)}(t, \alpha)= (1-t/T)(H_{D}+\alpha H_{XX})+(t/T)H_{P}^{(XXZ)}$. We plot the energy diagram when the problem Hamiltonian is the XXZ model, as shown in Fig. \ref{fig:xxz_with_non-stoquastic_energy}. There is a level crossing; hence, the ground-state search with QA does not succeed even when we consider an infinitely long annealing time.
We discuss this point using an analytical method (see Appendix \ref{sec:appendix_symm}).

\begin{figure}[ht]
    \includegraphics[width=80mm]{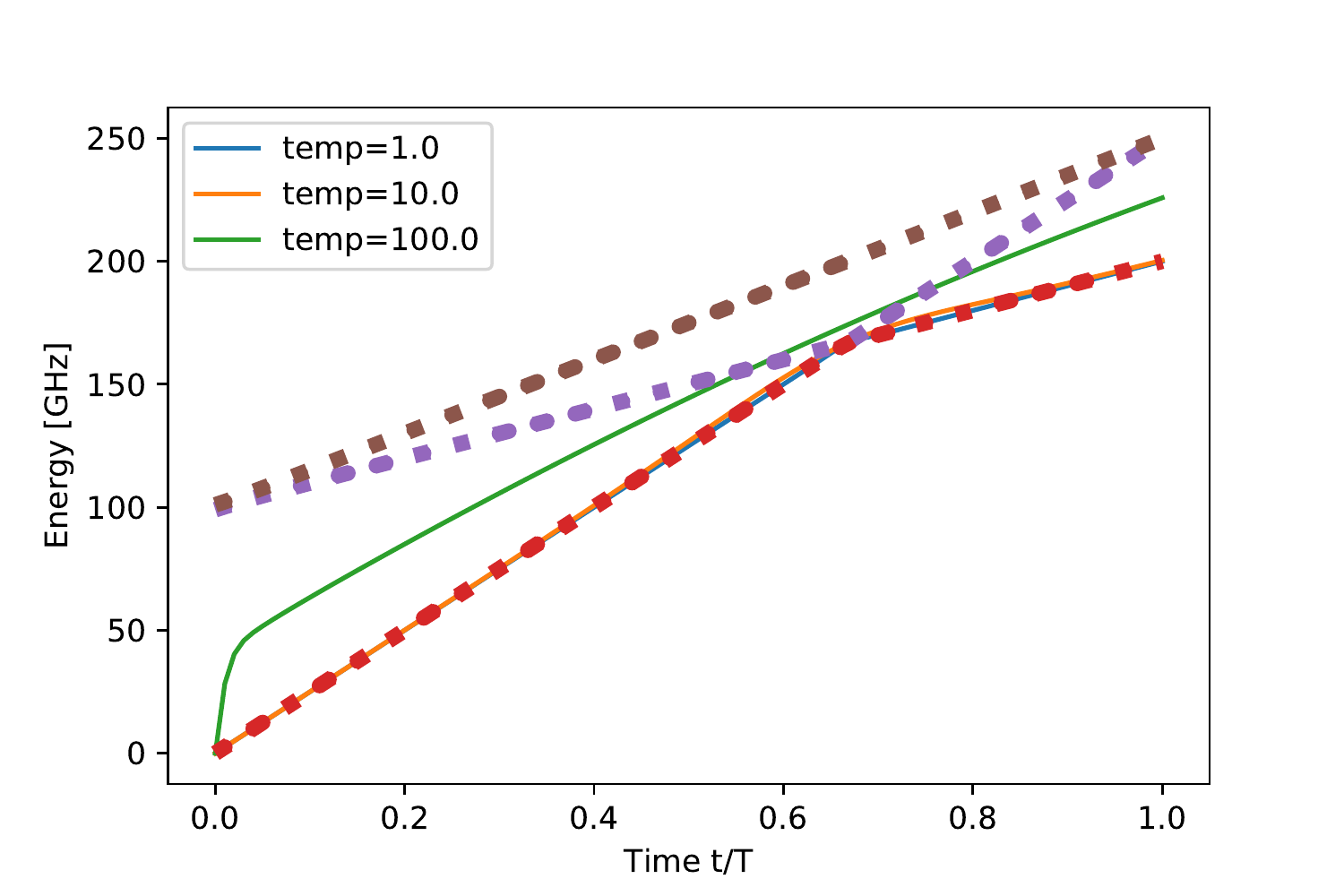}
    \caption{
    With the XXZ model as the problem Hamiltonian,
    we plot the energy against $t/T$ via QA for the temperatures of $1.0, 10.0, 100.0$ by using continuous lines.
    The dotted line represents the energy spectrum against each time $t/T$.
    QA succeeds in preparing a ground state with high accuracy if the temperature of the environment is sufficiently low.
    We set $\alpha=100$ and $T=1000$.
    }
    \label{fig:xxz_with_non-stoquastic_energy_decoherence}
\end{figure}

Similarly, we consider a noise operator to break the symmetry of the Hamiltonian.
We conduct the numerical calculation using the quantum adiabatic Markovian master equations.
As shown in Fig. \ref{fig:xxz_with_non-stoquastic_energy_decoherence},
we plot the energy spectrum and the energy expectation value
$\langle H\rangle ={\rm{Tr}}[H(t)\rho(t)]$ during QA.
In contrast to the case with unitary dynamics without noise, we succeed in obtaining the ground state
via QA with noise, especially when the environmental temperature is low.
The population of the excited state increases with the temperature owing to the thermal excitation.

\section{conclusion}

In this paper, we presented examples showing that, when a non-stoquastic Hamiltonian is used, catastrophic failure of QA occurs in the sense that we cannot obtain a ground state even with an infinitely long annealing time. Moreover, we find that we can avoid such failure by using a certain type of decoherence.
The key aspect of this finding is the symmetry of the Hamiltonian.
As there exists an observable to commute with the driver Hamiltonian and problem Hamiltonian, we can block-diagonalize the annealing Hamiltonian. When we add the anti-ferromagnetic interaction term (i.e., a typical non-stoquastic term) to the transverse-field driver Hamiltonian, the sector of the ground state of the problem Hamiltonian becomes different from that of the driver Hamiltonian. In this case, there is no transition from the ground state of the driver Hamiltonian to that of the problem Hamiltonian; hence, the ground-state search with QA does not succeed despite a long annealing time.
However, by adding decoherence to break the symmetry, the transition between the ground state of the driver Hamiltonian and that of the problem Hamiltonian becomes possible, and we can find the ground state using QA.
In summary, we challenged the common wisdom that a non-stoquastic Hamiltonian improves the performance of QA whereas decoherence degrades the performance. Therefore, our results provide a deeper understanding of QA.

\begin{acknowledgments}

This work was supported by Leading Initiative for Excellent Young Researchers MEXT Japan and JST presto
(Grant No. JPMJPR1919) Japan. This paper is partly
based on results obtained from a project, JPNP16007,
commissioned by the New Energy and Industrial Technology Development Organization (NEDO), Japan.
\end{acknowledgments}

\appendix
\section{Analytical derivation of symmetry}\label{sec:appendix_symm}
In this section, we analytically investigate the symmetry of the annealing Hamiltonian.
As the state is confined in the maximum angular momentum, we use the Dicke basis for our analysis.

\subsection{Fully connected Ising model}
We consider the fully connected Ising model.
In this case, the drive Hamiltonian and the problem Hamiltonian are given by
\begin{align}
    H_{D}&=B_{x}\hat{S}_{x}+\chi \hat{S}_{x}^{2}\\
    H_{P}&=D_{0}\hat{S}_{z}^{2}.
\end{align}
We assume that $|B_{x}|\ll\chi$.
The ground state of $H_{P}$ is $\ket{S_{z}=0}$.
Here, a parity operator is denoted by $\hat{P}(=e^{i\pi\hat{S}_{x}})$.
We can derive the following properties: $\hat{P}\hat{S}_{x}\hat{P}=\hat{S}_{x}$, $\hat{P}\hat{S}_{y}\hat{P}=-\hat{S}_{y}$, and $\hat{P}\hat{S}_{z}\hat{P}=-\hat{S}_{z}$.
When we set $\chi=0$, the ground state of the driver Hamiltonian $H_{D}$ is $\ket{S_{x}=-j}$.
Meanwhile, if $\chi>B_{x}$, the ground state of the driver Hamiltonian $H_{D}$ is $\ket{S_{x}=0}$.

We can calculate the parity of these states as follows.
The parity of the ground state of the driver Hamiltonian
with or without the non-stoquastic term is given by

\begin{align}
    \hat{P}\ket{S_{x}=-j}&=(-1)^j\ket{S_{x}=-j}\\
    \hat{P}\ket{S_{x}=0}&=\ket{S_{x}=0}
\end{align}

where $j$ is an integer.
Meanwhile, the parity of the ground state for the problem Hamiltonian is obtained as follows:

\begin{align}
    \hat{P}\ket{S_{z}=0}=(-1)^j\ket{S_{z}=0}\label{eq:PSz=0}
\end{align}
Here, we use $\hat{P}(=e^{i\pi\hat{S}_{x}})=e^{i\frac{1}{2}\pi\hat{\sigma}^{(1)}_{x}} e^{i\frac{1}{2}\pi\hat{\sigma}^{(2)}_{x}} \cdots e^{i\frac{1}{2}\pi\hat{\sigma}^{(2j)}_{x}}$ and $e^{i\frac{1}{2}\pi\hat{\sigma}_{x}}|\downarrow\rangle =-i |\uparrow\rangle $ where $|\uparrow\rangle $ and $|\downarrow\rangle $ are eigenstates of $\hat{\sigma}_z$.

Therefore, we show that if $j$ is odd and $\chi$ is sufficiently large, the sector of the ground state of the driver Hamiltonian is different from that of the problem Hamiltonian.

\subsection{Fully connected XXZ model}

Here, we analyze the symmetry of the fully connected XXZ model.
The Hamiltonian of the fully connected XXZ model is given by
\begin{align}
    H_{P}=S_{x}^{2}+S_{y}^{2}+\Delta S_{z}^{2}
\end{align}
where $\Delta$ denotes the anisotropic parameter.
If the anisotropic parameter $\Delta$ is sufficiently large, the ground state of this Hamiltonian is $\ket{S_{z}=0}$, which is the same as that explained in the Appendix \ref{sec:appendix_symm}.
Since the driver Hamiltonian is the same as that of the Appendix \ref{sec:appendix_symm}, the parity of the ground state is also the same as that of the Appendix \ref{sec:appendix_symm}. Therefore, we show that the sector of the ground state of the driver Hamiltonian is different from that of the problem Hamiltonian for large $\chi$ and $\Delta$ when $j$ is odd.

\section{Numerical simulation using the GKSL master equation}\label{sec:numerical_gksl}

In this subsection, we consider numerical simulation using the GKSL master equation instead of the quantum adiabatic Markovian master equations without the rotating wave approximation.
In the main text, we described calculations conducted using the quantum adiabatic Markovian master equations 
at low temperatures, and we showed that we can successfully prepare the ground state with noise. This is because the noise operator contains information of the system Hamiltonian, which actually induces a transition from the excited state to the ground state of the system.
Meanwhile, in the GKSL master equation, which is used frequently, the noise operator (or Lindblad operator), which does not contain information of the system Hamiltonian with interactions, is phenomenologically selected. 
Therefore, the ground state of the system cannot be prepared by noise as long as we employ the GKSL master equation.
In this section, we quantitatively investigate this point by conducting numerical simulations.

Typically, the GKSL master equation in a finite-temperature bath is given by
\begin{align}
    \frac{d}{dt}\rho(t)=&\eta(N+1)\biggl[\hat{L}\rho(t)\hat{L}^{\dag}-\frac{1}{2}\{\hat{L}^{\dag}\hat{L}, \rho(t)\}\biggr]\notag\\
    &+\eta N\biggl[\hat{L}^{\dag}\rho(t)\hat{L}-\frac{1}{2}\{\hat{L}\hat{L}^{\dag}, \rho(t)\}\biggr]
\end{align}
where $N=\frac{1}{e^{1/T}-1}$, $\hat{L}=\sum_{j=1}^{N}\hat{\sigma}_{-}$ and $T$ is the temperature.
It is worth mentioning that, although this type of GKSL master equation is widely used, this noise operator can be derived from a microscopic model when non-interacting qubits are considered. In the case of interacting qubits, we should consider a stricter model, such as quantum adiabatic Markovian master equations. 

\begin{figure}[ht]
    \includegraphics[width=80mm]{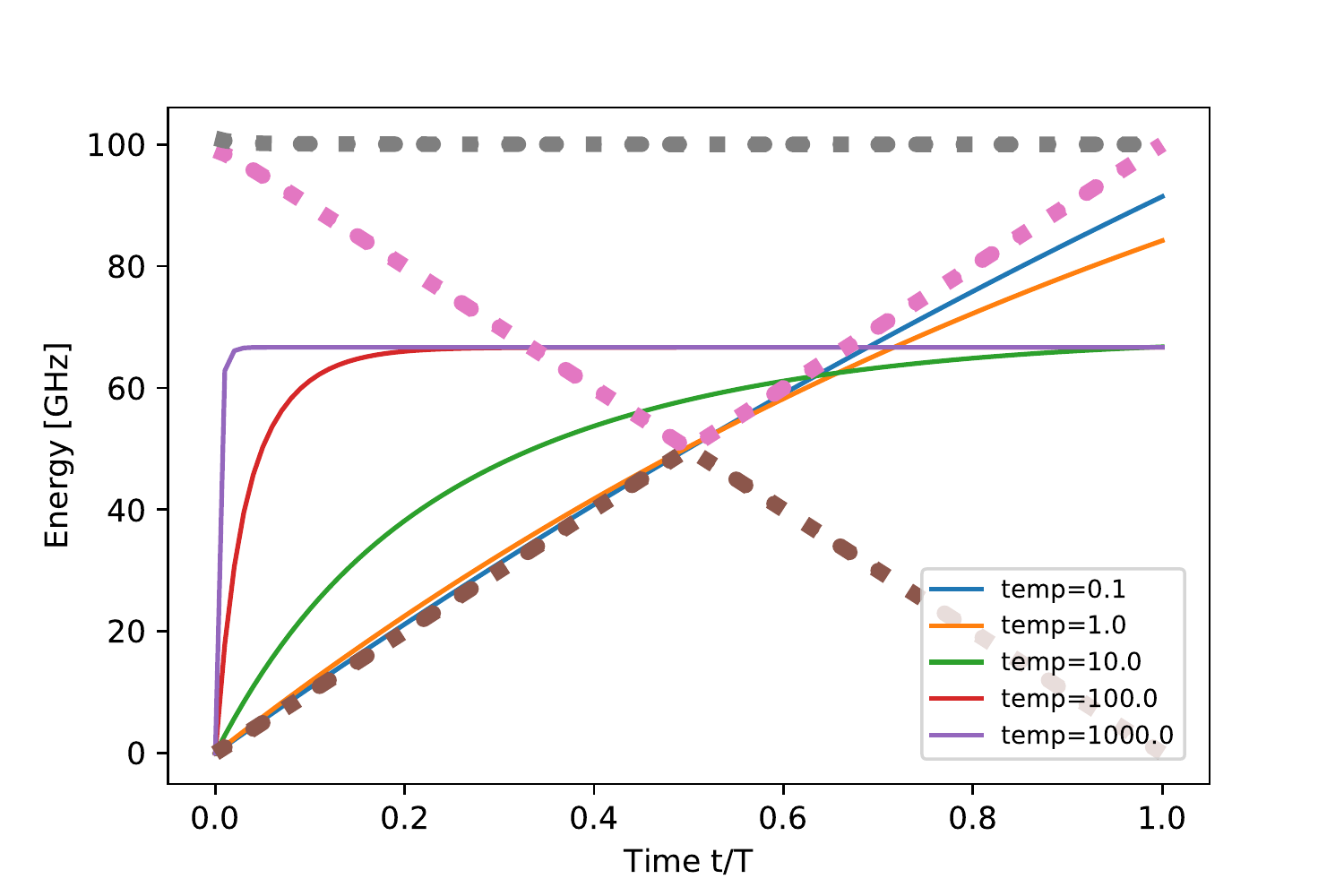}
    \caption{With the fully connected Ising model as the problem Hamiltonian,
    we plot the energy expectation obtained by QA using the GKSL master equation against each time $t/T$.
    We set $\alpha=100$, $\eta=0.1$, and $T_{env}=0.1, 1.0, 10.0, 100.0, 1000.0$.
    The dotted line represents the energy spectrum against each time $t/T$.
    We confirm that we cannot prepare the ground state with QA under the effect of decoherence described by the GKSL master equation.
    }
    \label{fig:ising_no_non-stoquastic_energy_gksl}
\end{figure}

\begin{figure}[ht]
    \includegraphics[width=80mm]{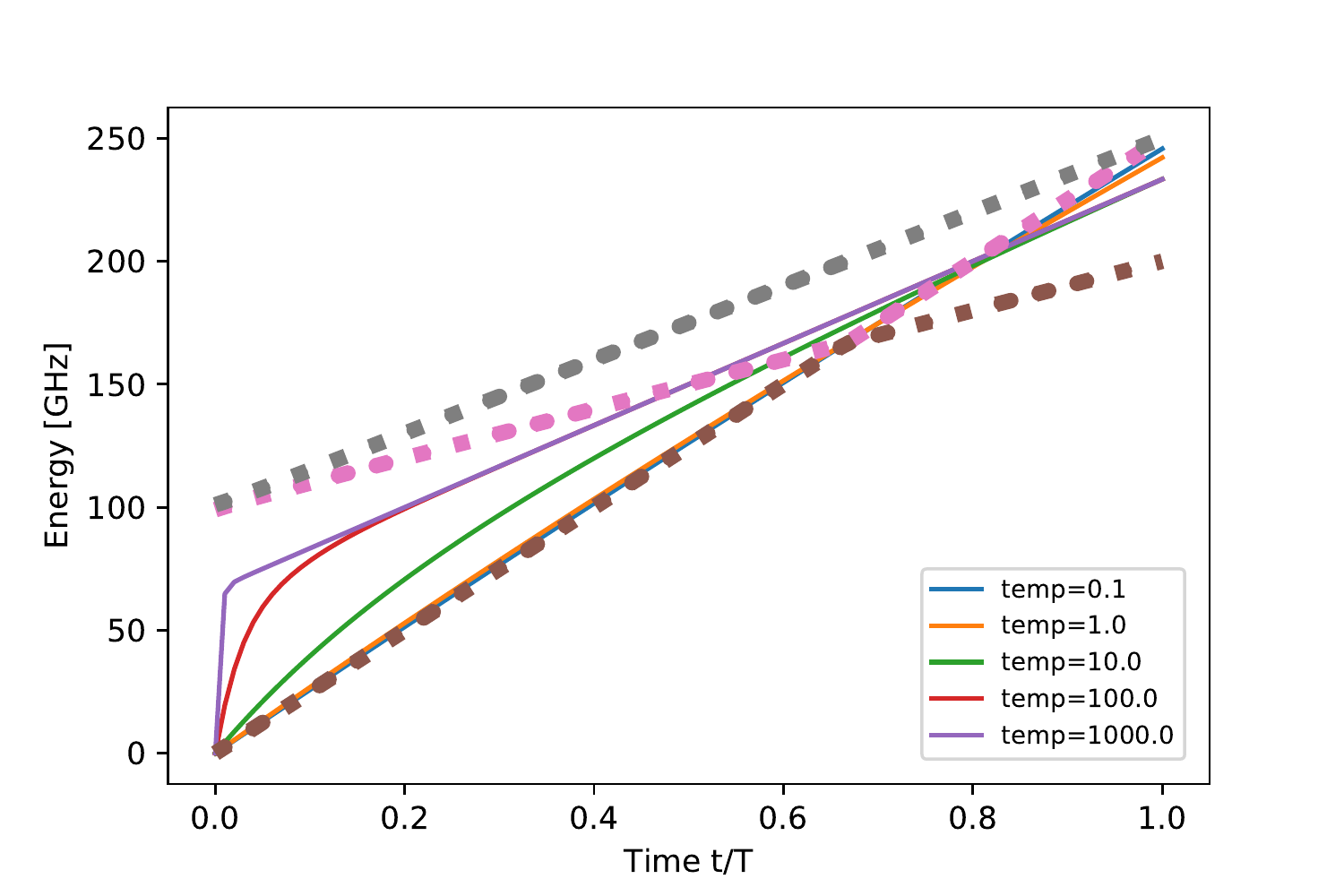}
    \caption{With the fully connected XXZ model as the problem Hamiltonian,
    we plot the energy expectation obtained by QA using the GKSL master equation against each time $t/T$.
    We set $\alpha=100$, $T=1000$, $\eta=0.1$, and $T_{env}=0.1, 1.0, 10.0, 100.0, 1000.0$.
    The dotted line represents the energy spectrum against each time $t/T$.
    We confirm that we cannot prepare the ground state with QA under the effect of decoherence described by the GKSL master equation.
    }
    \label{fig:xxz_no_non-stoquastic_energy_gksl}
\end{figure}

First, we consider the fully connected Ising model.
Let us consider the annealing Hamiltonian with a non-stoquastic term $(1-t/T)(H_{D}+\alpha H_{XX})+(t/T)H_{P}^{(Ising)}$.
Here, $H_{P}^{(Ising)}$ is described in Eq.\ref{nononsp}, $H_{D}$ is described in Eq.\ref{nononsd}, and $H_{XX}$ is described in Eq.\ref{eq:xx_non-stoquastic}.
We set $\alpha=100$, $\eta=0.1$, and 
$T_{env}=1.0, 10.0, 100.0$.
We conduct the numerical calculation using the GKSL master equation for this annealing Hamiltonian.
We plot the energy spectrum and the energy expectation value, as shown in Fig. \ref{fig:ising_no_non-stoquastic_energy_gksl}.
This figure shows that the fidelity between the ground state and the state obtained by QA does not increase as the temperature decreases.
At high temperatures, the state obtained by QA is a maximally complete mixed state.
By contrast, at low temperatures, the state obtained by QA is the all-down state.
Therefore, we show that, by using the GKSL master equation to describe noise during QA, we cannot prepare the ground state.
This highlights the importance of using an accurate noise model derived from a microscopic model such as quantum adiabatic Markovian master equations.

Next, we consider the fully connected XXZ model.
Let us consider the annealing Hamiltonian with a non-stoquastic term $(1-t/T)(H_{D}+\alpha H_{XX})+(t/T)H_{P}^{(XXZ)}$, where $H_{P}^{(XXZ)}$ is given by Eq. \ref{eq:xxz}.
We set $\alpha=100$, $\eta=0.1$, and 
$T_{env}=10.0, 1000.0$.
Furthermore, we plot the energy expectation obtained by the numerical simulation of QA using the GKSL master equation (in Fig. \ref{fig:xxz_no_non-stoquastic_energy_gksl}).
As with the fully connected Ising model, decoherence is not useful for preparing the ground state
if we use the GKSL equation with phenomenologically selected Lindblad operators.

\section{Numerical simulations
in the case of many qubits
}\label{manyqubits}

\begin{figure}[h]
    \includegraphics[width=80mm]{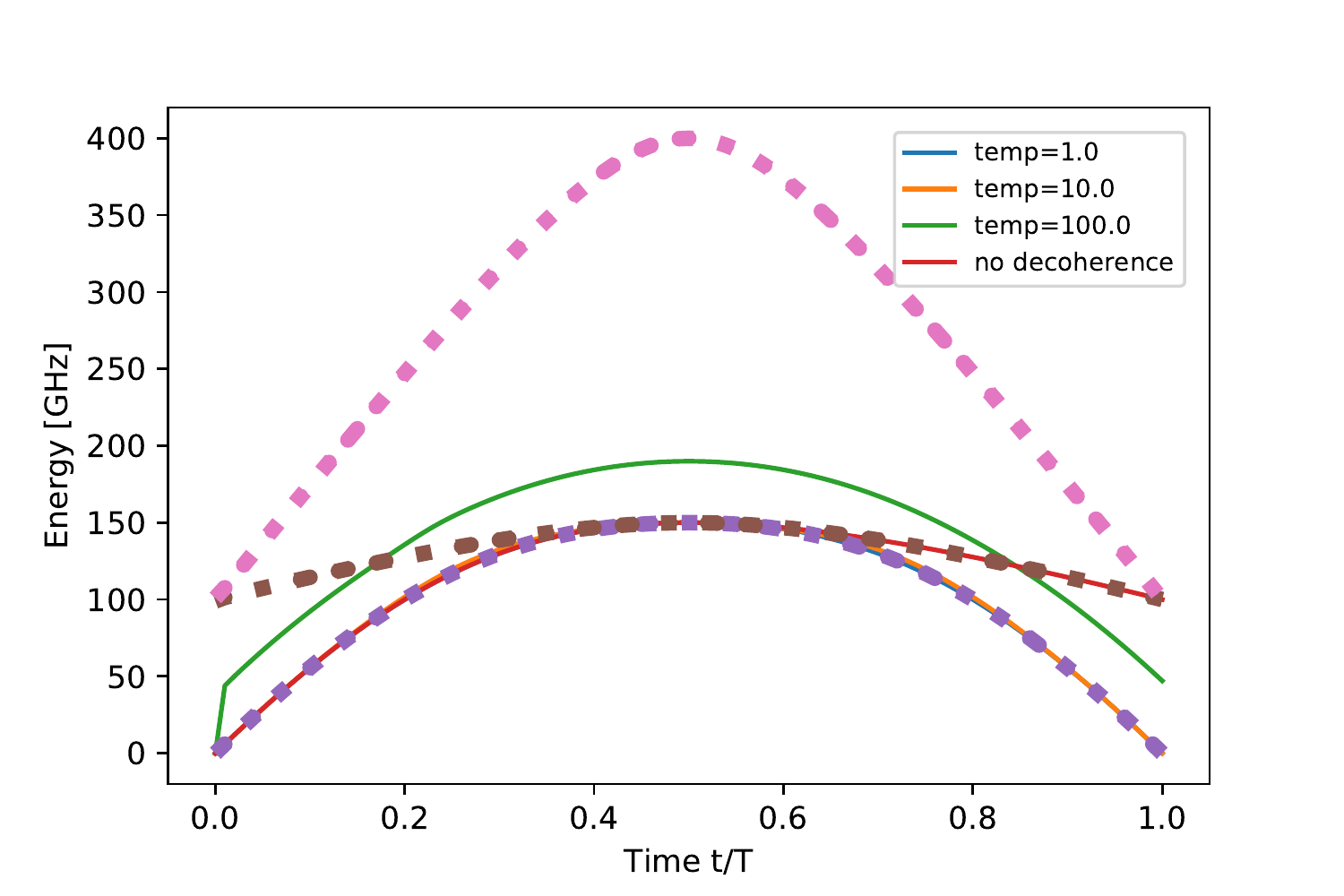}
    \caption{With the fully connected Ising model as the problem Hamiltonian and $S=3$ (7 qubits),
    we plot the energy expectation obtained by QA using the GKSL master equation against each time $t/T$.
    We set $\alpha=100$, $T=1000$, $\eta=0.1$, and 
    $T_{env}=1.0, 10.0, 100.0$.
    In addition, we plot the no-decoherence case at $\alpha=100$, $T=1000$.
    The dotted line represents the energy spectrum against each time $t/T$.
    }
    \label{fig:ising_no_non-stoquastic_energy_s=3}
\end{figure}

\begin{figure}[h]
    \includegraphics[width=80mm]{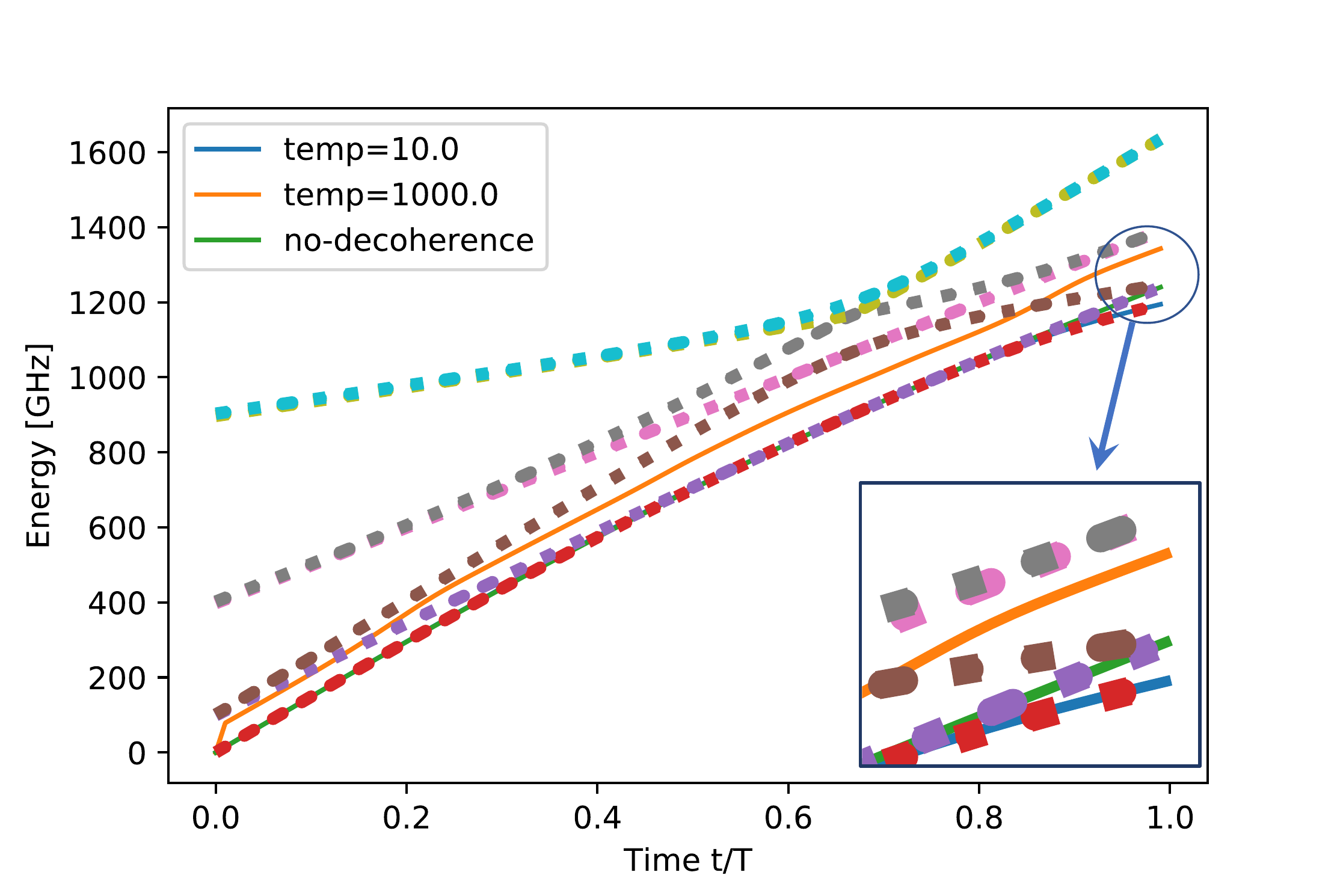}
    \caption{With the fully connected XXZ model as the problem Hamiltonian and $S=3$ (7 qubits)
    at $\Delta=1.5$,
    we plot the energy expectation obtained by QA using the GKSL master equation against each time $t/T$.
    We set $\alpha=100$, $T=1000$, $\eta=0.1$, and 
    $T_{env}=10.0, 1000.0$.
    In addition, we plot the no-decoherence case at $\alpha=100$, $T=1000$.
    The dotted line represents the energy spectrum against each time $t/T$.
    }
    \label{fig:xxz_no_non-stoquastic_energy_s=3}
\end{figure}

In the main text, we described numerical simulations conducted with $S=1$ (2 qubits).
In this section, we investigate the case of increasing the number of qubits.
More specifically, we consider the case of $S=3$ (7 qubits).
As shown in Fig. \ref{fig:ising_no_non-stoquastic_energy_s=3} and Fig. \ref{fig:xxz_no_non-stoquastic_energy_s=3}, we plot the energy spectrum and the energy expectation value $\braket{H} = Tr[H(t)\rho(t)]$ during QA for the fully connected Ising model and the fully connected XXZ model.
From these, we can see that the method adopted in this paper works well even when the number of spins increases.

\newpage


\bibliographystyle{apsrev4-1}
\bibliography{apssamp}

\end{document}